\documentclass{article}

\usepackage{arxiv}

\usepackage[utf8]{inputenc} 
\usepackage[%
colorlinks=true,%
linkcolor=blue,%
citecolor=blue,%
urlcolor=blue]%
{hyperref}       
\usepackage{url}            
\usepackage{booktabs}       
\usepackage{amsmath}
\usepackage{amssymb}
\usepackage{amsfonts}       
\usepackage{nicefrac}       
\usepackage{graphicx}
\usepackage{natbib}
\usepackage{doi}
\usepackage{booktabs}
\usepackage{appendix}
\usepackage{wrapfig}

\renewcommand{\vec}[1]{\mathbf{#1}}

\title{Influence of passenger head-position uncertainty on infection risk predictions}
\date{\today}

\newif\ifuniqueAffiliation

\ifuniqueAffiliation 
\author{ Philipp Bahavar\thanks{philipp.bahavar@dlr.de}, Florian Webner, Andrei Shishkin, and Daniel Schmeling \\
	German Aerospace Center\\
	Institute of Aerodynamics and Flow Technology\\
	37073 Göttingen, Germany
}
\else
\usepackage{authblk}

\author[a]{%
	Philipp~Bahavar\thanks{{Corresponding author: philipp.bahavar@dlr.de}}}%
\author[a,b]{%
	Florian~Webner}%
\author[a]{%
	Andrei~Shishkin}%
\author[a]{%
	Daniel~Schmeling}%
\affil[a]{German Aerospace Center, Institute of Aerodynamics and Flow Technology, 37073 Göttingen, Germany}
\affil[b]{Institute of Thermodynamics and Fluid Mechanics, Technische Universität Ilmenau, 98693 Ilmenau, Germany}
\fi

\renewcommand{\headeright}{}
\renewcommand{\undertitle}{}
\renewcommand{\shorttitle}{Head-position uncertainty and infection risk predictions}
\hypersetup{
pdftitle={Influence of passenger head-position uncertainty on infection risk predictions},
pdfsubject={physics.flu-dyn, q-bio.QM},
pdfauthor={Philipp Bahavar},
pdfkeywords={CFD, Covid, UQ, PCE},
}

\begin{document}
\maketitle

\begin{abstract}
We investigate the influence of natural head movement on the infection risk posed by airborne pathogens using a CFD-based forward risk prediction model.
Quasi-Monte-Carlo simulations are used to obtain the resulting infection risk distributions by representing head movement via probability distributions of parameters describing the position and orientation of each passenger's breathing zone.
A significant impact of fore/aft and lateral head position on infection risk was found and should be accounted for to increase robustness when predictions of local, seat-specific infection risks are used to guide design and policy decisions.
Unlike the sampling-based Monte Carlo approach, estimates of the statistical moments of the risk distributions and sensitivities, calculated using first-order second-moment and higher-order methods, were found to inadequately capture the dependencies between head movement components and infection risk.
\end{abstract}

\section{Introduction}
\label{sec:introduction}
The risk of spreading infections on public transport has become a matter of increased focus in the wake of the COVID-19 pandemic.
Airborne transmission in particular has been identified as the dominant mode of infection with SARS-CoV-2 in the enclosed spaces of passenger cabins and compartments \citep{Zhang2020, Qian2020}.

Going forward, aerosol spreading and the resulting infection risk caused by airborne transmission will need to be considered when designing cabin layouts and ventilation concepts as well as when making policy decisions regarding occupancy limitations.
Robust computational methods will be required to predict these risks \citep{Azzini2020}.
Such models must include methods to take into account and quantify the uncertainty of the model results caused by the input parameter uncertainties, as previously established in various applications, including the oil and gas industry \citep{Bickel2008}, environmental regulations \citep{Frey1992}, and public health applications in the context of epidemics and pandemics at the population level, both before and after the COVID-19 pandemic \citep{Gilbert2014,Taghizadeh2020,Swallow2022}.

In both experimental and numerical investigations of aerosol spreading performed in aircraft cabins \citep{Schmeling2023,Shishkin2023}, spatially inhomogeneous particle concentrations are found as a consequence of the flow patterns that emerge from the interplay of the cabin geometry, the ventilation setup, and convective flow contributions from the heating caused by the passengers themselves.
Therefore, models aiming to predict the infection risk for each passenger must reach beyond the assumption of a well-mixed room \citep{Wells1955,Riley1978} to yield accurate results.

When performing calculations using an infection risk model capable of resolving the aforementioned spatially inhomogeneous particle distributions such as described by \cite{Webner2024} and considered in the present investigation, both ``frozen flow'' and ``frozen passengers'' are customarily assumed.
This eliminates transient effects from consideration, thereby significantly reducing the computational demands.
Consequently, the particle transport calculations only consider the average flow fields, and the evaluation of which fraction of the surrounding particle cloud falls into the passenger-specific breathing zone and ultimately contributes to their infection risk only considers the average position of each passenger.

This study focuses on the second assumption, investigating how the natural head movements of the passengers during a long-distance train journey introduce positional uncertainty into the model, influencing the predicted infection risk.
To this end, the head movements are reduced to variable positions and orientations of the breathing zones, which in turn are no longer fixed to specific values, but are instead defined by probability distributions with specific statistical properties.
Performing Monte Carlo simulations by sampling from these distributions and subsequently evaluating the infection risk model for the set of sampled inputs generates the corresponding infection risk distribution.
These risk distributions are then used to evaluate the accuracy of estimates obtained directly from the properties of the input distributions using local, derivative-based methods.
Additionally, global variance-based sensitivity analysis is performed using polynomial chaos expansion to extract the sensitivities of the predicted risk with respect to the different components of passenger head movement, thus identifying the leading contributions to risk variance.

Starting in section \ref{sec:methods}, the infection risk prediction toolchain, consisting of the flow simulation and particle transport on the one hand and the infection risk model on the other hand, is outlined.
Section \ref{sec:uq} then discusses the methods used for uncertainty quantification and sensitivity analysis, which are subsequently applied in section \ref{sec:results} to the prediction of the average infection risks for all occupants of a long-distance train compartment caused by a single passenger exhaling infectious SARS-CoV-2 pathogens.
After comparing the different approaches, section \ref{sec:conclusion} draws conclusions on the application of uncertainty and sensitivity analysis to capture the effects of the natural head movements of the train passengers.

\section{Risk prediction methodology}
\label{sec:methods}
The numerical prediction of the average infection risks is based on two pillars.
Firstly, computational fluid dynamics (CFD) simulations of the flow field and of the aerosol particle transport provide the evolution of the particle concentration throughout the geometry.
Based on these results, the infection model then translates the particle concentration and age within a breathing zone into the associated individual infection risk in a second step.

\subsection{CFD approach to aerosol spreading}
\label{subsec:methods:cfd}
The spread of exhaled aerosol particles throughout the train compartment is computed based on the average flow fields obtained through CFD simulations.
The simulation setup follows the practices established in previous work on aerosol spreading in an aircraft cabin \citep{Schmeling2023,Shishkin2023}.
For this purpose, the interior geometry is decomposed into a computational mesh using the meshing tools available in OpenFOAM \citep{OpenFoam2025}.
The base cell size is $20\,\mathrm{mm}$, with refinement down to $5\,\mathrm{mm}$ towards the surfaces.
The average flow is then calculated using a \emph{Reynolds-averaged Navier--Stokes} (RANS) simulation using the $k\!-\!\omega-$SST turbulence model \citep{Menter1994, Menter2003}.
Second-order accurate discretization schemes are employed, with upwind schemes used for the convective terms of the transport equations and central schemes for the remaining terms.
The specific solver used is \texttt{buoyantBoussinesqSimpleFoam}, which includes heat transfer and the linear approximation for the buoyancy force in flows with limited density variations \citep{Gray1976} that can otherwise be treated as incompressible.
Thermal convection is primarily driven by the passengers, who are represented by thermal manikins in the simulations.
Their heat flux varies by body part, depending on the skin temperature and clothing coverage.
Therefore, temperature differences exist between the passengers and the inflow air entering the compartment through the ventilation outlets, as well as between the passengers and the outer walls of the geometry, which are exposed to the ambient temperature.

The resulting average flow field is then used as a frozen representation of the flow conditions that drive the spreading of aerosol particles.
Spherical particle clouds with zero initial velocity are seeded in front of the passengers' ``faces''.
The particle sizes are uniformly distributed between $0.5\,\mathrm{\mu m}$ -- $5\,\mathrm{\mu m}$, corresponding to the particle sizes relevant for spreading the SARS-CoV-2 virus \citep{Alsved2022}.

Lagrangian particle transport simulations are then performed for each of the initial aerosol cloud positions using the \texttt{icoUncoupledKinematicParcelFoam} solver.
The aerodynamic drag forces acting on a spherical particle \citep{Putnam1961} are given by
\begin{align}
	F_{d}=\frac{3}{4}\frac{\rho \nu C_{d}\mathit{Re}_{p}}{\rho_{p}d_{p}^{2}},\label{eq:sphere_drag_force}
\end{align}
where $\rho$ is the density and $\nu$ is the kinematic viscosity of the fluid, $\rho_{p}$ the is density and $d_{p}$ is the diameter of the particle.
The particle Reynolds number is
\begin{align}
	\mathit{Re}_{p}=\frac{u_{\mathit{rel}}d_{p}}{\nu},\label{eq:particle_Reynolds}
\end{align}
where $u_{\mathit{rel}}$ is the magnitude of the relative velocity between the particle and the collocated fluid element.
Finally, the drag coefficient is \citep{Amsden1989}
\begin{align}
	C_{d} = \begin{cases} \frac{24}{\mathit{Re}_{p}}\left(1 + \frac{1}{6}\mathit{Re}_{p}^{\frac{2}{3}}\right),\quad &\text{if }\mathit{Re}_{p} \leq 1000,\\
	0.424,\quad &\text{if }\mathit{Re}_{p} > 1000.\label{eq:sphere_drag_coeff}
\end{cases}
\end{align}

Only the forces acting on the particles are computed, while the reaction forces acting on the fluid are disregarded (\emph{one-way coupling}), which is justified for the low particle concentrations considered here \citep{Elghobashi1994}.

To account for the variations in trajectory and additional particle dispersion caused by the interaction between particles and turbulent eddies, an additional stochastic dispersion is added based on the local turbulent kinetic energy $E_{k}$ using the \texttt{stochasticDispersionRAS} model \citep{Talebmoustaph2024}.
In this approach, the fluid velocity acting on the particle is modified by adding a fluctuation
\begin{align}
	\vec{u}^{\prime} = \left(\frac{2E_{k}}{3} \right)^{\frac{1}{2}}\,\vec{r},\label{eq:stochastic_dispersion}
\end{align}
where $\vec{r}$ is a random vector with uniformly distributed orientation and normally distributed magnitude.

Using this numerical setup, the spread of the aerosol particles is simulated for a total time of $1000\,\mathrm{s}$, with one simulation performed for each initial particle cloud position.
The resulting particle positions are recorded every second of simulated time, providing the input for the infection risk prediction model.

\subsection{Infection risk modeling}
\label{subsec:methods:infection}
A two-step risk model is applied to predict the passenger-specific infection risk \citep{Webner2024}.
First, the inhaled infectious dose of a susceptible individual is estimated based on the CFD simulation results.
The infection probability is then calculated in the second step using a dose--response relationship derived from the SARS-CoV-2 human challenge study  \citep{Killingley2022}.

Only the air volume in the direct proximity to the passenger is relevant during inhalation.
The risk prediction model accounts for this by establishing breathing zones -- hemispheres with a radius of $0.2\,\mathrm{m}$ placed in front of and aligned with the passengers' heads \citep{OSHA}, described by the center position of the hemisphere and two angles specifying its orientation.
The pathogen concentrations within this zone then determine the local, individual infection risk following the Wells--Riley equation \citep{Ko2004,Gao2008}.

The inhaled infectious dose $D_{i}$ for a specific passenger denoted by the index $i$ is then proportional to the number of active viral payloads present in the corresponding breathing zone during the time frame $t$.
This number, in turn, depends on the intensity of virus emission at the source, given by the emission rate $R$, which determines the total number of infectious agents released into the compartment per unit time, and is given in units of the \emph{tissue culture infectious dose} (TCID$_{50}$).

Transport into the relevant breathing zone as computed by the Lagrangian particle simulation with seeding rate $\dot{P}$ results in a zone-averaged particle concentration $C_{i}$.
Since the particle concentration $C_{i}$ is approximately proportional to the particle seeding rate $\dot{P}$, the ratio $C_{i}/\dot{P}$ is proportional to the fraction of emitted particles that reach the breathing zone.
This ratio is independent of the chosen seeding rate. However, a higher seeding rate may reduce statistical noise and improve accuracy, albeit at the cost of increased computational effort.
Taking into account the particle-specific transport time $\tau_{p}$ and consequently the decay of active virus over time via the survival fraction $s_{i}(\tau_{p})$, the resulting exposure of the passenger in question then results from multiplying by the pulmonary ventilation rate $p_{\mathit{in}}$.
In summary, the inhaled infectious dose $D_{i}$, again expressed in multiples of TCID$_{50}$, is calculated as
\begin{align}
\label{eq:inhaled_dose}
D_{i} = \frac{R s_{i}C_{i}}{\dot{P}} \,p_{in} t.
\end{align}
The fraction $s_{i}$ of virus in breathing zone $i$ that remains active during transport is computed as the average over all particles in a breathing zone,
\begin{align}
\label{eq:average_survival_fraction}
s = \frac{1}{n_{p}}\sum^{n_{p}}_{p=1} s(\tau_{p}),
\end{align}
where $n_{p}$ is the number of particles in the breathing zone, $s(\tau)$ is the airborne stability function of the virus over time.
Here, exponential decay at a rate of $0.6\,\%$ per minute is assumed \citep{Dabisch2020}.
The chosen values of all parameters are summarized in Table \ref{tab:parameter_values}.

\begin{table}[h!]
\label{tab:parameter_values}
\centering
\caption{Typical ranges for the physiological and medical model parameters and specific values chosen for this investigation.}
\begin{tabular}{llrrl}
\toprule
Parameter & Unit & Possible range & Chosen value & Reference \\
\midrule
emission rate $R$ & TCID$_{50} / \mathrm{s}$ & $<5.5$ & $0.05$ & \citep{Zheng2022,Lai2022} \\
TCID$_{50}$ & - & $10^{3}$ -- $10^{6}$ & $10^{4}$ & \citep{Sender2021} \\
ventilation rate $p_{\mathit{in}}$ & $\mathrm{L/s}$ & $0.1$ -- $0.9$ & $0.1$ & \citep{Hinds2022} \\
exposure time $t$ & h & 0.5 -- 10 & 2 & --- \\
\bottomrule
\end{tabular}
\end{table}

To estimate the dose--response relationship, we assume that each inhaled infectious virus poses the same probability to cause infection \citep{Watanabe2010}.
Based on this assumption, an exponential dose--response curve is fitted to the human challenge data \citep{Killingley2022}, in which 18 of 34 volunteers became infected ($I_{\mathrm{HCD}}=0.53$) after inoculation with a dose $D_{\mathrm{HCD}}=10\, \text{TCID}_{50}$.
The resulting dose--response function,
\begin{align}
\label{eq:infection_risk}
I(D_{i}) = 1 - \exp{\left[\log\left(\frac{1-I_{\mathrm{HCD}}}{D_{\mathrm{HCD}}}\right)\,D_{i}\right]} = 1 - \exp{(-0.0754\, D_{i})},
\end{align}
gives the infection risk $I$ as a function of the inhaled dose as calculated using equations \ref{eq:inhaled_dose} and \ref{eq:average_survival_fraction}.

\section{Quantification of risk uncertainty and sensitivity}
\label{sec:uq}

\subsection{Local derivative-based approximations}
\label{subsec:uq:local}
For determining the uncertainty of the predicted infection risks and for analyzing their sensitivity on the input parameters of the infection model, we consider $k$ inputs $x_i\in\left\lbrace x_{1},x_{2},\ldots x_{k}\right\rbrace$.
Each $x_{i}$ is an independent random variable drawn from a distribution $X_{i}$, which is characterized by an expected value
\begin{align}
	E\left[X_{i}\right]=\mu_{i}\label{eq:Xi_expectation}
\end{align}
and finite variance
\begin{align}
	V\left[X_{i}\right] = \sigma_{i}^{2} < \infty.\label{eq:Xi_variance}
\end{align}

Let now $Y=f\left(\vec{x}\right)$ be the result of the infection model risk calculation for the vector of input parameters $\vec{x}=\left(x_{1},\ldots,x_{k}\right)$.
A robust prediction of the infection risk associated with a specific geometry or flow setup consequently needs to provide accurate estimates of both $E\left[Y\right]=\mu_{Y}$ and $V\left[Y\right]=\sigma_{Y}^{2}$.

The first approach for estimating these quantities investigated here is the \emph{first-order second moment} method (FOSM) \citep{Elishakoff1987}.
This method utilizes the Taylor expansion of $f$ about the point $\vec{x}_{\mu}=\left(\mu_{1},\ldots \mu_{k}\right)$ to obtain \citep{Kriegesmann2012}
\begin{align}
	\mu_{Y}\approx f\left(\vec{x}_{\mu}\right)\label{eq:FOSM_mean}
\end{align}
and
\begin{align}
	\sigma_{Y}^{2}\approx \sum_{i=1}^{k}\left.\frac{\partial Y}{\partial x_{i}}\right\rvert_{\mu_{i}}^{2} \sigma_{i}^{2}.\label{eq:FOSM_variance}
\end{align}

Since the derivatives of the model output with respect to the inputs are not known a priori, they are computed numerically using a central differencing scheme,
\begin{align}
	\left.\frac{\partial Y}{\partial x_{i}}\right\rvert_{\mu_{i}} = \frac{f\left(\vec{x}_{\mu}+\Delta \vec{x}_{i}\right) - f\left(\vec{x}_{\mu}-\Delta \vec{x}_{i}\right)}{2\Delta x_{i}},\label{eq:central_1st_derivative}
\end{align}
where $\Delta \vec{x}_{i}$ is a vector whose $i$-th components is equal to $\Delta x_{i}$ and all other components are zero.
Thus, the FOSM needs a total of $2k+1$ evaluations of the infection risk model to provide estimated means and variances.

Due to its nature as a first-order approximation, the FOSM is not well-suited for nonlinear dependencies.
If the model's response to an input $x_{i}$ is nonlinear, the approximation of the first derivative, given by equation (\ref{eq:central_1st_derivative}), does not generalize to the entire range of inputs covered by the distribution $X_{i}$.
Consequently, the estimated variance then depends on the choice of $\Delta x_{i}$.
Additionally, the approximation of the mean output from (\ref{eq:FOSM_mean})  becomes inaccurate due to the missing higher-order effects of the input.
To address these challenges, higher-order terms of the Taylor expansion can be used together with higher statistical moments of the input distributions to calculate corrections to the first-order estimates.
For example, the \emph{second-order third moment method} (SOTM) estimates the mean of the output as
\begin{align}
	\mu_{Y}\approx f\left(\vec{x}_{\mu}\right) + \frac{1}{2} \sum_{i=1}^{k} \left.\frac{\partial^{2} Y}{\partial x_{i}^{2}}\right\rvert_{\mu_{i}} \sigma_{i}^{2},\label{eq:SOTM_mean}
\end{align}
and the variance as
\begin{align}
	\sigma_{Y}^{2} =\sum_{i=1}^{k}\left.\frac{\partial Y}{\partial x_{i}}\right\rvert_{\mu_{i}}^{2} \sigma_{i}^{2} + \left(f\left(\vec{x}_{\mu}\right)\right)^{2} -\mu_{Y}^{2}f\left(\vec{x}_{\mu}\right)\sum_{i=1}^{k}\left.\frac{\partial^{2} Y}{\partial x_{i}^{2}}\right\rvert_{\mu_{i}}\sigma_{i}^{2} + \sum_{i=1}^{k} \mu_{3,i} \left.\frac{\partial Y}{\partial x_{i}}\right\rvert_{\mu_{i}}\left.\frac{\partial^{2} Y}{\partial x_{i}^{2}}\right\rvert_{\mu_{i}},\label{eq:SOTM_variance}
\end{align}
where $\mu_{3,i}$ is the third-order central moment of $X_{i}$ \citep{Kriegesmann2012}.
Computing the necessary second derivatives via
\begin{align}
	\left.\frac{\partial^{2} Y}{\partial x_{i}^{2}}\right\rvert_{\mu_{i}} = \frac{f\left(\vec{x}_{\mu}+\Delta\vec{x}_{i}\right) - 2\,f\left(\vec{x}_{\mu}\right) + f\left(\vec{x}_{\mu}-\Delta\vec{x}_{i}\right)}{\Delta x_{i}^{2}}\label{eq:central_2nd_derivative}
\end{align}
means no additional model evaluations are necessary compared to the FOSM, but the skewness of the input distributions needs to be known in addition to the mean and variance as assumed above.

Although higher-order methods are better suited to account for nonlinear effects, the estimated values for the mean and variance of the output remain sensitive to the choice of $\Delta x_{i}$, particularly if $f$ depends irregularly on $x_{i}$ and changes qualitatively across the range of input values.

The question of the \emph{sensitivity} of the output to certain inputs can be restated as the question of what fraction of the output variance is caused by the variance of each input.
This question is immediately answered by the FOSM and higher order methods.
Dividing equation (\ref{eq:FOSM_variance}) by $\sigma_{Y}^{2}$ and looking at each component of the sum individually,
\begin{align}
	S_{\sigma_{i}} = \frac{\sigma_{i}^{2}}{\sigma_{Y}^{2}} \left.\frac{\partial Y}{\partial x_{i}}\right\rvert_{\mu_{i}},\label{eq:sigma_norm_derivatives}
\end{align}
yields the \emph{sigma-normalized derivatives} \citep{Penman2000} or \emph{importance factors} \citep{ASME_VV20_2009}, an established measure of model sensitivity.
Analogous treatment of the higher order estimate formulated in equation (\ref{eq:SOTM_variance}) results in a similar decomposition of the total output variance into contributions of the individual inputs.
Assuming the derivative-based approach is suitable overall, these normalized derivatives $S_{\sigma_{i}}$ quantify the relative sensitivity of the overall infection risk to a specific input $i$ and allow comparisons between the set of inputs as a whole.

\subsection{Quasi-Monte Carlo simulations and global sensitivity analysis}
\label{subsec:uq:qmc}
The methods outlined in the previous section aim to estimate the mean and variance of the model output based on the local value of the derivatives, necessitating only a limited number of model evaluations.
In contrast, \emph{Monte Carlo} methods (MC) aim to directly obtain the distribution of the model output.
To this end, the distributions of the inputs $X_{i}$ are sampled at random to generate input vectors $\vec{x}_{n}$ and the corresponding model output $Y_{n}=f(\vec{x}_n)$.
Repeating this process for $N$ samples causes the observed distribution of all $Y_{n}$ to converge for sufficiently large $N$, allowing the calculation of the statistical moments of interest.

Using MC to obtain the mean risk and its variance circumvents the challenges posed by the potentially nonlinear or irregular dependence on the model inputs.
Sampling the entire distribution and evaluating the model at these points captures any complex effects on the output distribution.

The convergence rate of MC is typically of order $\mathcal{O}(\sqrt{N})$.
This rate can be improved by transitioning from randomly sampling the inputs to pseudo-random sampling according to low-discrepancy sequences.
This leads to a \emph{quasi-Monte-Carlo} simulation (QMC) \citep{Niederreiter1978}, with improved convergence rate of the order $\mathcal{O}(\log(N)^{k}/N)$, reducing the number of model evaluations necessary to obtain reliable values for the mean and variance.
Here, QMC is used as implemented in the Python package \texttt{SciPy} \citep{SciPy2020}, using a Sobol' sequence \citep{Sobol1967} with scrambling \citep{Owen1998}.
The specific structure of the pseudo-random sequence restricts the total number of samples to powers of two ($N=2^{m}$) to preserve the improved convergence characteristics \citep{Owen2020}.

While obtaining the output distribution and its statistical properties is straightforward (if potentially expensive) using (Q)MC, the decomposition of the variance into separate contributions becomes more challenging compared to the local methods and their normalized derivatives.
To address this problem, consider first the Sobol' decomposition of $f$ itself,
\begin{align}
	f\left(\vec{x}\right) = f_{0} + \sum_{i=1}^{k} f_{i} + \sum_{i=1}^{k}\sum_{j=i+1}^{k} f_{ij} + \ldots + f_{1,2,\ldots,k},\label{eq:HDMR_decomposition}
\end{align}
forming a high-dimensional model representation (HDMR) \citep{Rabitz1999} with
\begin{align}
	f_{0} =& E\left[Y\right],\notag \\
	f_{i} =& E\left[Y\rvert x_{i}\right] - f_{0},\label{eq:HDMR_coeffs}\\
	f_{ij} =& E\left[Y\rvert x_{i}, x_{j}\right] - E\left[Y\rvert x_{i}\right] - E\left[Y\rvert x_{j}\right] - f_{0}\notag \\
	\vdots \notag
\end{align}
providing the best fit to $Y$ \citep{Ratto2007}.
Here, $E[Y\rvert \lbrace x_{i}\rbrace]$ refers to the conditional expectation value of $Y$ for a set of fixed inputs $\lbrace x_{i}\rbrace$.
From this, the variance $V[Y]$ can be similarly decomposed into
\begin{align}
	V\left[Y\right] = \sum_{i=1}^{k}V_{i} + \sum_{i=1}^{k}\sum_{j=i+1}^{k} V_{ij} + \ldots + V_{1,2,\ldots,k},\label{eq:variance_decomposition}
\end{align}
where
\begin{align}
	V_{i} =& V\left[E\left[Y\rvert x_{i}\right]\right],\label{eq:variance_coeffs}\\
	\vdots\notag
\end{align}
analogous to the definitions in (\ref{eq:HDMR_coeffs}).
Dividing by the total variance $V$ then yields the sensitivity indices $S$,
\begin{align}
	1 = \sum_{i=1}^{k}S_{i} + \sum_{i=1}^{k}\sum_{j=i+1}^{k} S_{ij} + \ldots + S_{1,2,\ldots,k}.\label{eq:sensitivities_all}
\end{align}
As the number of summands in equations (\ref{eq:HDMR_decomposition}) -- (\ref{eq:sensitivities_all}) grows as $2^{k}-1$ with the number of inputs, sensitivity analysis focuses on two indices in particular:
The \emph{first-order sensitivity indices} $S_{i}$ (also \emph{first-order Sobol' indices}) that contain the effect on the output variance of $x_{i}$ alone (the \emph{main effect}), and the \emph{total sensitivity indices} (\emph{total Sobol' indices}) $S_{T_{i}}$, which is the sum over all terms in equation (\ref{eq:sensitivities_all}) that contain the index $i$ \citep{Saltelli2010} and quantify the \emph{total effect} of the input in question.

Various estimators exist for both $S_{i}$ and $S_{T_{i}}$ based on specific schemes of mutiple MCs with partially overlapping input samples, necessitating $(k+2)N$ model evaluations in total \citep{Sobol1993,Homma1996,Jansen1999,Saltelli2010}.

A different approach to extracting the sensitivity indices from QMC directly utilizes the decomposition of the model given by equation (\ref{eq:HDMR_decomposition}).
A polynomial of fixed order is fitted to the set of model outputs $Y_{1},\ldots,Y_{N}$ for the inputs $\vec{x}_{1},\ldots,\vec{x}_{N}$, yielding the coefficients $c_{\alpha}$ of the \emph{polynomial chaos expansion} (PCE):
\begin{align}
	Y \approx Z\left(\vec{x}\right)=\sum_{\alpha} c_{\alpha} \psi_{\alpha}(\vec{x}),\label{eq:PCE}
\end{align}
where the $\psi_{\alpha}$ are a set of polynomial basis functions \citep{Sudret2008}.
By choosing the basis such that the functions are orthogonal with respect to the input distributions $X_{i}$,
\begin{align}
	\left\langle \psi_{\alpha},\psi_{\beta}\right\rangle \equiv \int_{X_{i}}\! \psi_{\alpha}\psi_{\beta}\;\mathrm{d}\vec{x} = \delta_{\alpha\beta},\label{eq:orthogonality}
\end{align}
the decomposition of the variance from equation (\ref{eq:variance_decomposition}) corresponds directly to
\begin{align}
	V\left[Y\right] \approx V\left[Z\right] = \sum_{\alpha} c_{\alpha}^{2},\label{eq:PCE_variance_decomposition}
\end{align}
where the index $\alpha$ corresponds to specific index combinations $ij\ldots$ \citep{Crestaux2009}.
Consequently, after fitting the PCE to the output, all sensitivity indices $S_{ij\ldots}$ are available analytically.
$S_{i}$ and $S_{T_{i}}$ specifically are obtained by summing over the $c_{\alpha}^{2}$ where $\alpha$ refers to polynomials containing powers of $x_i$ exclusively or at all, respectively.

Because the PCE-based approach accounts for the complete output variance present in the empirical sample provided by QMC, the Sobol' indices obtained are global in the sense that they express the effect of the different inputs across the whole range of the input distributions, instead of being constrained to a small interval around the anchor point of the derivative-based local approaches \citep{Saltelli2010}.
By providing a sufficient number of QMC samples to fit a PCE with an appropriately chosen maximum degree $d$ -- minimizing the mean squared error of the fit across the dataset while avoiding overfitting -- yields Sobol' indices that quantify and compare the $k$ different model inputs, determining their relative importance \citep{Sobol2001}.

\section{Infection risk analysis for long-distance passenger train compartment}
\label{sec:results}
\begin{figure}[ht]
    \centering
    \includegraphics[width=\linewidth]{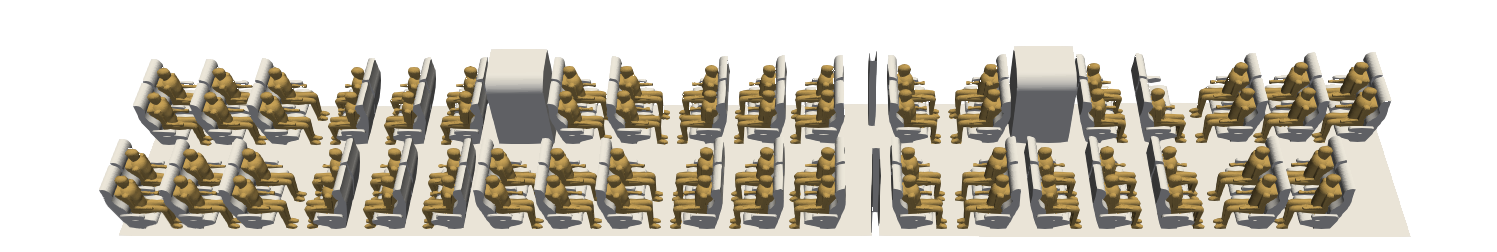}
    \caption{Overview of the long-distance train compartment, including dividers, luggage racks and 73 passengers seated both row-wise and vis-à-vis.}
    \label{fig:cabin_layout}
\end{figure}

Using the methodology outlined in section \ref{subsec:methods:cfd}, the average flow fields were computed for a long-distance passenger train compartment occupied by 73 seated passengers.
Then, the Lagrangian aerosol spreading simulations were performed for sources corresponding to the exhalation of each passenger.
The geometry represents a compartment of the first and the second generation of the German Inter City Express (ICE) long-distance passenger train.
As shown in figure \ref{fig:cabin_layout}, the compartment layout includes both front- and backwards-facing seats, passengers seated across from each other at a table, and luggage storage racks, thus providing a variety of different local passenger--passenger constellations.
The ventilation outlets are placed along the length of the train overhead of the aisle, and the cabin air exhausts along the side walls beneath the seats.
The total airflow rate is $3700\,\mathrm{m^{3}/h}$, corresponding to a supply airflow rate of $50\,\mathrm{m^{3}/h}$ per seat.
Inflow air temperature is set to $T_{\mathit{in}}=18\,\mathrm{^{\circ}C}$.
An ambient temperature of $T_{\mathit{amb}}=26\,\mathrm{^{\circ}C}$ is applied to the exterior of the outer walls, which have a thermal transmittance of $U_{\mathit{wall}}=1\,\mathrm{W/(m^{2}K)}$.
The passenger heat flux varies based on the body-part-specific skin temperatures ranging from $34.75\,\mathrm{^{\circ}C}$ to $35.85\,\mathrm{^{\circ}C}$ and clothing coverage representing light summer clothing with a clothing value of $0.5\,\mathrm{clo}$ ($U_{\mathit{cloth}}=12.9\,\mathrm{W/(m^{2}K)}$).

To generate a single representative risk value for each passenger, all possible source positions are taken into consideration by evaluating the infection risk model using the different particle spreading results obtained by the source-specific Lagrangian simulations.
Then, by averaging over the individual risks for each possible source (excluding cases where the source and receiver are identical), an expectation value for the seat-specific infection risk is obtained.
This expected risk then serves as the quantity of interest in the analysis of the influence of motion-induced position uncertainty of the breathing zones.

\begin{figure}[ht]
    \centering
    \includegraphics[width=0.65\linewidth]{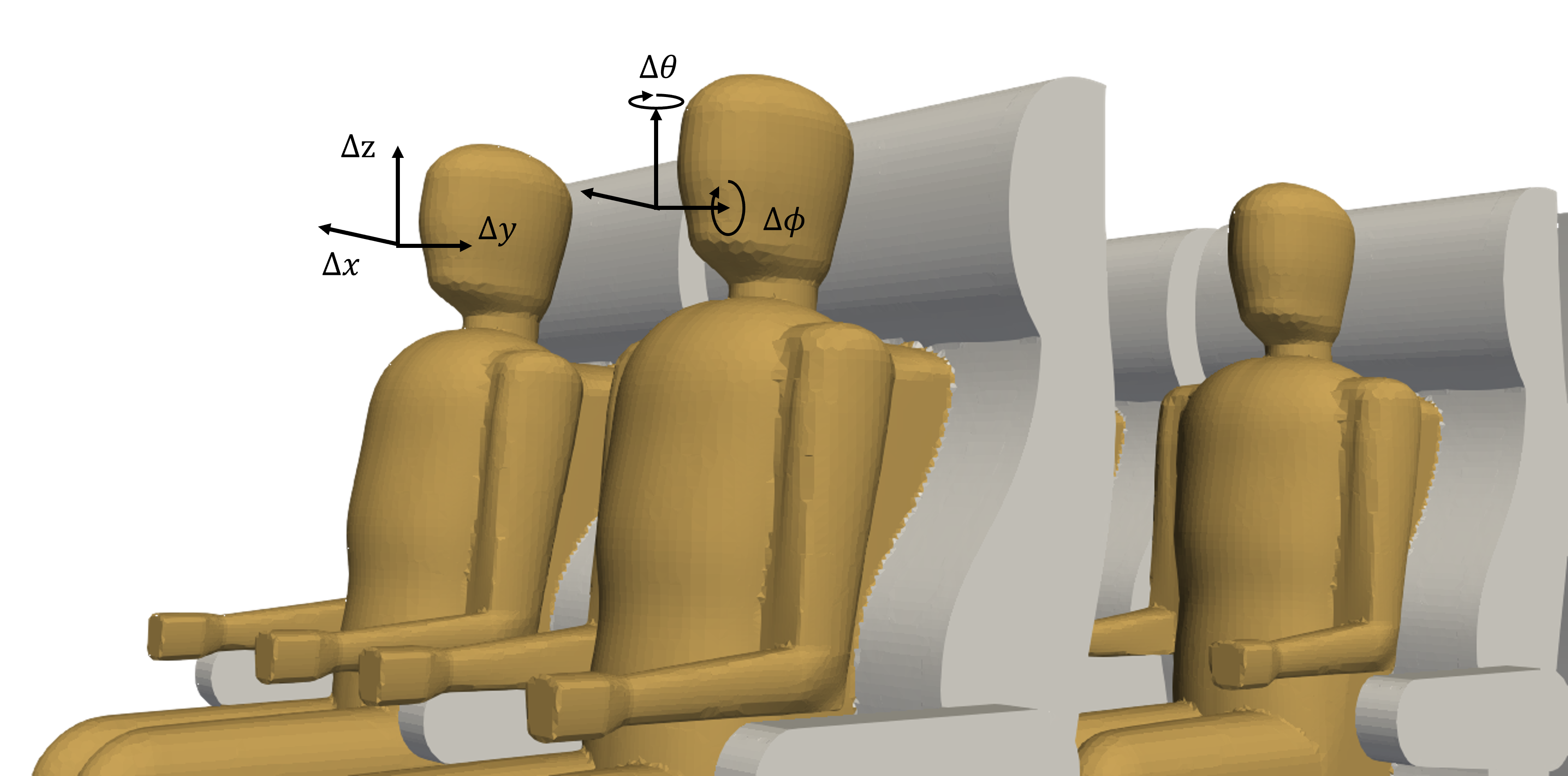}
    \caption{Illustration of the passenger coordinate system in which the offsets of the breathing zones are defined. Investigated are linear shifts along the coordinate axes as well as angular offsets corresponding to rotating the breathing zones left--right ($\theta$) and up--down ($\phi$). }
    \label{fig:passenger_coords}
\end{figure}

Based on the particle distributions resulting from the Lagrangian simulations, QMC simulations are performed by sampling offsets along the three coordinate axes, as well as tilting ($\phi$) and turning angles ($\theta$) from normal distributions, modifying the idealized positions and orientations of the breathing zones.
Figure \ref{fig:passenger_coords} illustrates the coordinate system for these offsets in relation to the passenger head positions.
The varied parameters $x_{i}$ and their  means $\mu_{i}$ and standard deviations $\sigma_{i}$ characterizing the distributions are summarized in table \ref{table:QMC_parameters}.
Since the range of passenger head positions is constrained in the backwards direction by the headrests, the distribution governing the offset $\Delta x$ is shifted forward by $+\sigma_{x}$.
These values provide an idealized representation of the range of head positions covered by the natural motion of a seated passenger, which is assumed to be normally distributed around the mean position for simplicity.

\begin{table}[h]
	\caption{For QMC, the five parameters corresponding to linear offsets and turning and tilting of the breathing zones were sampled from the normal distributions $\mathcal{N}(\mu_{i},\sigma_{i})$ with means and standard deviations as shown here.}\label{table:QMC_parameters}
	\centering
	\begin{tabular}{lccccc}
		\toprule
		input parameter $x_{i}$ & $\Delta x$ & $\Delta y$ & $\Delta z$ & $\Delta \theta$ & $\Delta \phi$ \\
		mean $\mu_{i}$ & $0.1\,\mathrm{m}$ & $0$ & $0$ & $0$ & $0$\\
		standard deviation $\sigma_{i}$ & $0.1\,\mathrm{m}$ & $0.2\,\mathrm{m}$ & $0.05\,\mathrm{m}$ & $30^{\circ}$ & $5^{\circ}$\\
		\bottomrule
	\end{tabular}
\end{table}

\begin{figure}[h]
	\centering
	\includegraphics[width=\textwidth]{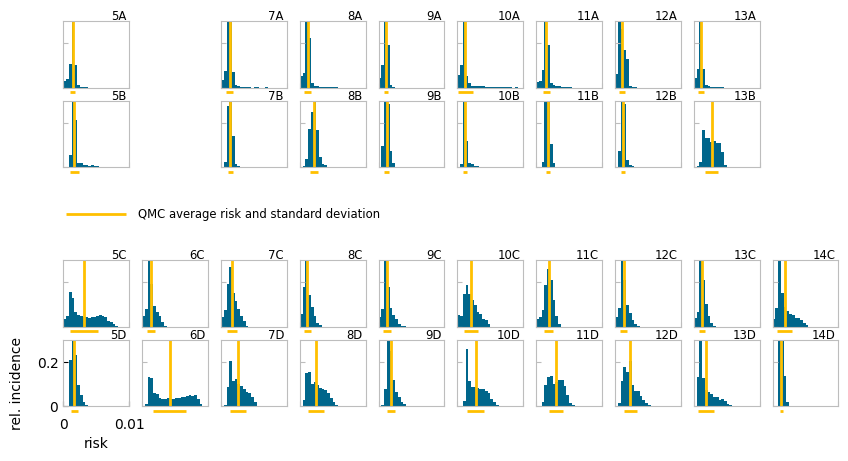}
	\caption{Histograms showing the distributions of the mean infection risk for the passengers seated in rows 5 through 14, averaged over all possible source positions, as produced by QMC simulations of $N=4096$ samples. The mean value and standard deviation for each seat are marked by the position of the vertical line within and the extent of the horizontal lines under the plots, respectively. }\label{fig:QMC_seat_histograms}
\end{figure}

The distributions of the average infection risk is obtained by calculating the average infection risk over all possible source locations for each sampled input vector.
Independent QMC runs using $N=2^{m},\,m=4,\,5,\ldots,\, 13$ were performed to ensure convergence of the output distributions before further analysis.
Results are shown here for $N=2^{12}=4096$.
Figure \ref{fig:QMC_seat_histograms} shows the risk distributions for passengers seated in rows 5 through 14.
The results for all seats are shown in appendix \ref{sec:app:full_results}, figure \ref{fig:seat_risks_full} for better readability.

The resulting risk distributions vary greatly depending on the specific seat.
Some distributions exhibit only weak dependence of the average risk on the varying inputs, resulting in sharply peaked distributions, as e.g. seen for passengers seated in row 9.
Other positions show a high variance of the risk, resulting in broad distributions, as seen for seats 5C and 6D, among others.
Overall, input combinations in which the predicted average risk differs from the average over the whole set of inputs by a factor of two in either direction are not uncommon.
This further strengthens the motivation to quantify the effects of the input variance on the risk prediction.

To investigate the feasibility of estimating the mean infection risks $\mu_{Y}$ alongside their standard deviations $\sigma_{Y}$ across input variations using local, derivative-based methods, the model was evaluated for the input vector corresponding to the mean of the inputs, $\vec{x}_{\mu}$ and for separate variations by $\Delta x_{i}=\pm \sigma_{i}$ along each of the five components.
These results were used to compute the first and second derivatives of the model response to each input using equations (\ref{eq:central_1st_derivative}) and (\ref{eq:central_2nd_derivative}).
Subsequently, $\mu_{Y}$ and $\sigma_{Y}$ were estimated via FOSM and SOTM.
A comparison of these results to those obtained from QMC is shown in figure \ref{fig:risk_comparisons}.
(Full results in appendix \ref{sec:app:full_results}, figure \ref{fig:seat_risks_full}.)

\begin{figure}
	\centering
	\includegraphics[width=\textwidth]{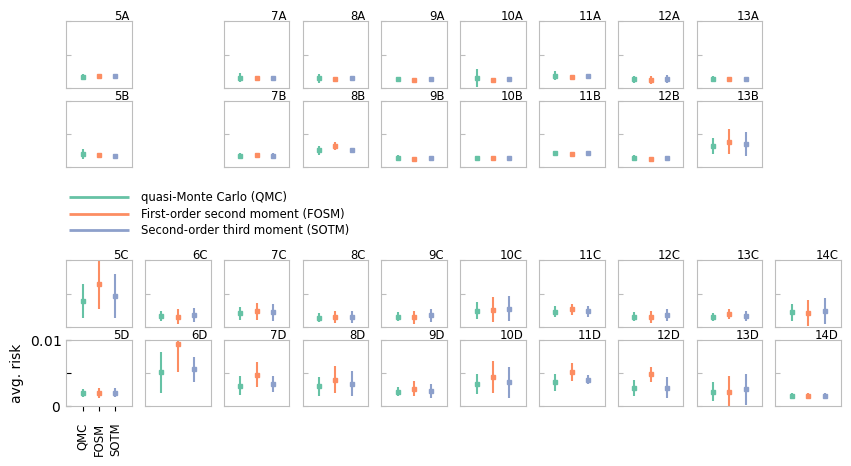}
	\caption{Infection risks for the passengers seated in rows 5 through 14, averaged over all possible source locations. Estimates for the mean $\mu_{Y}$ and standard deviation $\sigma_{Y}$ of the resulting risk distribution as computed by QMC, FOSM, and SOTM, shown using the data points and error bars, respectively. }\label{fig:risk_comparisons}
\end{figure}

While predictions for $\mu_{Y}$ agree for select seats, there is poor agreement between FOSM and QMC in aggregate over all passengers, with a root-mean-squared relative error (RMSRE) of $24.8\,\%$.
Using SOTM instead results in a significantly improved average risk prediction, with an RMSRE of only $9.4\,\%$.
This points towards nonlinear dependencies between the positional inputs and the infection risk, which are inadequately handled by the inherently linear FOSM, but captured by the higher-order approximation.
This is illustrated for example for seats 5C and 12D, where FOSM significantly overpredicts the average infection risk compared to QMC, while the SOTM result shows far better agreement

Notably, the tendency of the SOTM to improve upon FOSM in estimating the mean of the output distributions does not necessarily correspond to a better prediction of the output variance, as seen for example for seat 11D.
In this case, the SOTM provides a better mean risk estimate than FOSM, but the higher-order approach results in a worse prediction of the standard deviation than the first-order estimate.
Overall, comparing the predictions for $\sigma_{Y}$ to the QMC results shows an RMSRE of $49.9\,\%$ for FOSM and $51.2\,\%$ for SOTM, with no advantage for the higher-order method, despite the significant improvement observed in the estimated mean values.

In summary, the predicted statistical moments of the risk distributions obtained from FOSM are not suitable for robustly assessing of the influence of passenger head movements on the local infection risk.
Estimates using SOTM show good agreement with the results from QMC in terms of $\mu_{Y}$, however, the error in the estimated widths of the distributions is comparatively much larger.

The error values above quantify the overall accuracy of the risk prediction for individual passengers.
If instead the predicted infection risk of the passengers collectively is evaluated, the risks themselves are averaged over all passengers.
Comparing then between the prediction approaches naturally leads to lower discrepancies, since over- and underestimations now partially cancel out.
For this metric, there is better agreement between QMC and FOSM, with an error of $16.3\%$ compared to an RMSRE of $24.8\%$.
For SOTM, the error is reduced down to $6.9\%$ from an RMSRE of $9.4\%$.

To obtain the first-order and total sensitivity indices from the QMC results, a PCE is fitted to the output distributions of each seat.
The monic Hermite polynomials $\mathit{He}(x)$ are used as the orthogonal basis functions with respect to the normal distributions of the input parameters $x_{i}$ \citep{Sudret2008,Crestaux2009}.
After testing, a maximum polynomial degree of 6 was selected, which allowed for minimal regression error while avoiding overfitting of the dataset.
The sensitivity indices were then computed from the fit parameters by summing the squared coefficients of the appropriate polynomial terms, as outlined in section \ref{subsec:uq:qmc}.

\begin{figure}
	\centering
	\includegraphics[width=\textwidth]{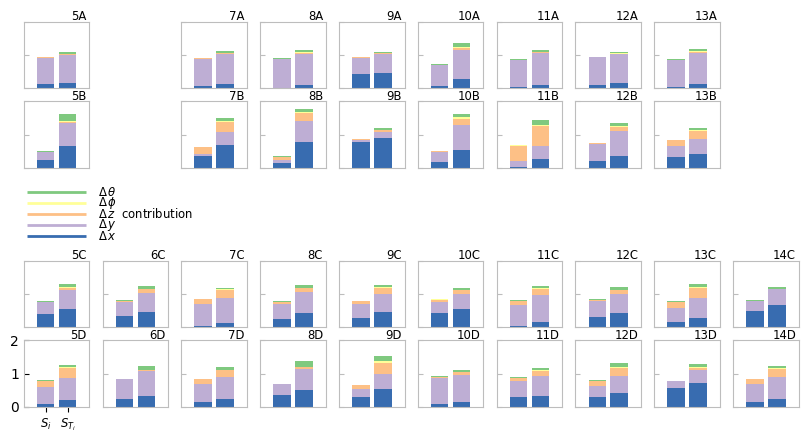}
	\caption{First-order ($S_{i}$, left columns) and total indices ($S_{T_{i}}$, right columns) of the five model inputs for the passengers seated in rows 5 through 14, computed from PCE of the QMC results. The respective height within the stacked bar corresponds to the value of the index in question.}\label{fig:first-total_comparison}
\end{figure}

Figure \ref{fig:first-total_comparison} shows the values of  both the first-order sensitivity indices $S_{i}$ and the total sensitivity indices $S_{T_{i}}$ for the five positional model parameters for passengers seated in rows 5 through 14.
(See appendix \ref{sec:app:full_results}, figure \ref{fig:sensitivities_full}, left panel for results for the complete compartment.)
The first-order and total indices both identify either the fore/aft offset $\Delta x$ or the lateral offset $\Delta y$ as the most sensitive parameters of the average infection risk.
These are followed by non-negligible contributions of the vertical offset $\Delta z$ to the total indices for a number of seats.
In contrast, the turning and tilting angles have virtually no direct effect as quantified by the first-order indices $S_{i}$.
However, $S_{T_{\theta}}$ in particular indicates sensitivities to rotation comparable to those of the vertical offset for specific seats, such as 5B or 9D.
This overall hierarchy of the sensitivities is also reflected in the average values taken over all seats as listed in table \ref{table:sensitivity_indices}.

\begin{table}[h]
	\caption{Average values of the first-order and total sensitivity indices for the average infection risk, taken over all seats.}\label{table:sensitivity_indices}
	\centering
	\begin{tabular}{lccccc}
		\toprule
		 & $\Delta x$ & $\Delta y$ & $\Delta z$ & $\Delta \theta$ & $\Delta \phi$ \\
		 first-order index $S_{i}$ & 0.215 & 0.479 & 0.059 & 0.015 & 0.002 \\
		 total index $S_{T_{i}}$ & 0.383 & 0.674 & 0.136 & 0.080 & 0.021 \\
		\bottomrule
	\end{tabular}
\end{table}

\begin{figure}
	\centering
	\includegraphics[width=\textwidth]{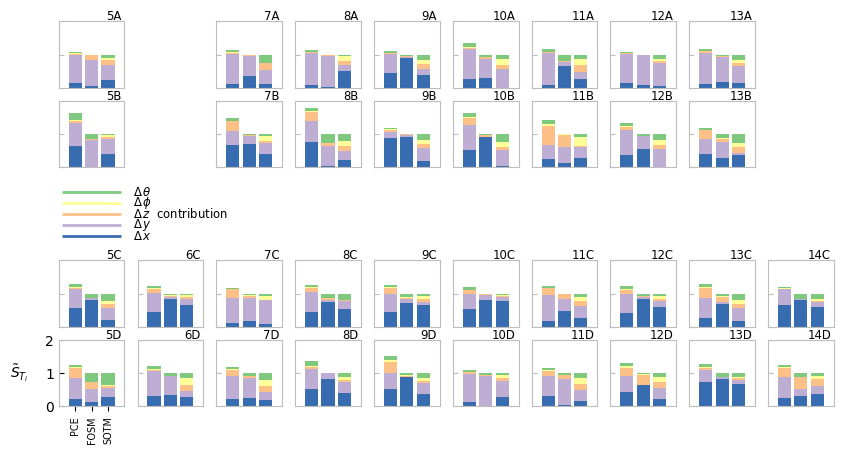}
	\caption{Total sensitivity indices $S_{T_{i}}$ of the position parameters for the passengers seated in rows 5 through 14 with respect to the infection risk averaged over all possible source locations as derived from PCE compared to sigma-normalized derivatives $S_{\sigma_{i}}$ obtained from FOSM and SOTM. }\label{fig:total_indices}
\end{figure}

The quality of the normalized derivatives for quantifying the sensitive parameters of the average infection risk is inconclusive when compared to the sensitivity indices derived from QMC via PCE.
The results for the total sensitivity indices $S_{T_{i}}$ are shown in figure \ref{fig:total_indices} compared to the sigma-normalized derivatives $S_{\sigma_{i}}$ computed from the contributions of the separate inputs to the variance predicted by FOSM and SOTM using equation (\ref{eq:sigma_norm_derivatives}).
The results for all passengers can be found in appendix \ref{sec:app:full_results}, figure \ref{fig:sensitivities_full}, right panel.

Considering only the identification of the most important contributing input from the hierarchy of  $S_{T_{i}}$ or $S_{\sigma_{i}}$, respectively, the local estimators agree with the PCE results for 41 out of 73 seats ($56\,\%$) for FOSM and 45 out of 73 seats ($62\,\%$) for SOTM.

While the hierarchies and even the relative magnitudes of the estimated sensitivities are accurate for some seats, overall agreement with the results of the global variance-based sensitivity analysis is poor, repeating the conclusions drawn above for the risk prediction of the variance itself using FOSM and SOTM.
Comparability is further hindered by the intrinsic normalization of the sigma-normalized derivatives, whereas the first-order and total indices can sum to values less than or greater than unity, respectively.

Similar to the qualitative differences in risk distributions obtained for each seat from the QMC, the results for the sensitivities highlight the variability in risk profiles for different positions within the compartment.

\begin{figure}
	\centering
	\includegraphics[width=0.6\textwidth]{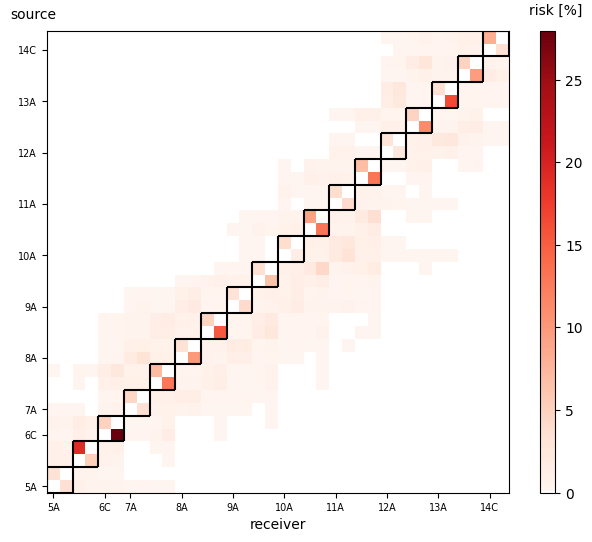}
	\caption{Average risk matrix obtained from QMC. Shown are the infection risks at a given receiver positions posed by a specific source positions for all possible combinations of seats in rows 5 through 14, with the risk magnitude represented by the color of the corresponding matrix entry. The black outlines group directly neighboring seats together.}\label{fig:risk_matrix}
\end{figure}

The high sensitivity of the average risk to offsets of the breathing zones in the $x$ and $y$ directions can be understood in the context of aerosol spreading in the cabin as follows.
Each source position contributes to the overall average of the infection risk at a given seat, however, the majority of the risk is caused by the few scenarios in which the source is located in close proximity, and especially if the source is in the directly neighboring seat.
Figure \ref{fig:risk_matrix} gives a visual representation of this fact, showing the risk posed by each source position to each receiver position.
Neighboring seats are grouped together by the outlines along the diagonal for visual guidance.
The risk matrix visualization for the complete compartment can again be found in appendix \ref{sec:app:full_results}, figure \ref{fig:risk_matrix_full}.
For all seats, the directly neighboring seat poses the highest average risk of infection.

\begin{figure}[htb]
    \centering
    \includegraphics[width=\linewidth]{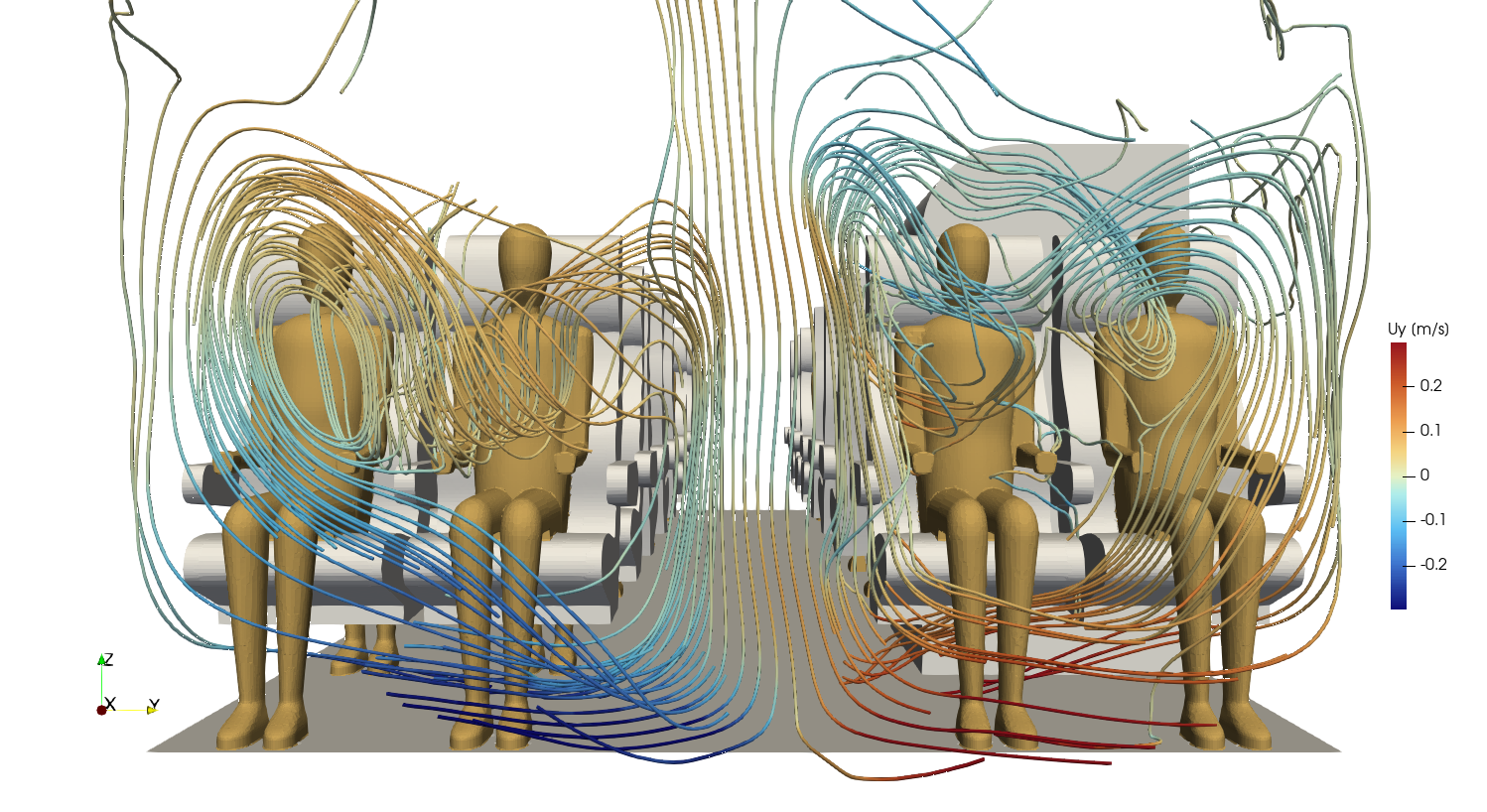}
    \caption{Streamlines of the flow between rows 8 and 9, showing the ventilation jet and the two counter-rotating circulations, resulting in flow from the window seat towards the aisle seat at head-height. Colors show the $y$-component of the flow velocity. }\label{fig:xnormal_streamlines_row8}
\end{figure}

This observation is explained by the flow topology, which shows row-wise organization of circulating flow in front of the passengers.
Figure \ref{fig:xnormal_streamlines_row8} illustrates the flow structures responsible for the importance of the direct neighbor for the overall infection risk by showing streamlines of the flow in the space between rows 8 and 9.
The ventilation setup with outlets in the ceiling above the central aisle produces a downward-facing ventilation jet that splits into two large-scale, counter-rotating circulations.
In these circulations, air flows along the floor and subsequently rises at the windows, flowing back towards the aisle at the height of the breathing zones.
This flow topology promotes the transport of aerosol particles between passengers seated directly next to each other, resulting in the block-diagonal structure of the risk matrix discussed above.

\begin{figure}
    \centering
    \includegraphics[width=\linewidth]{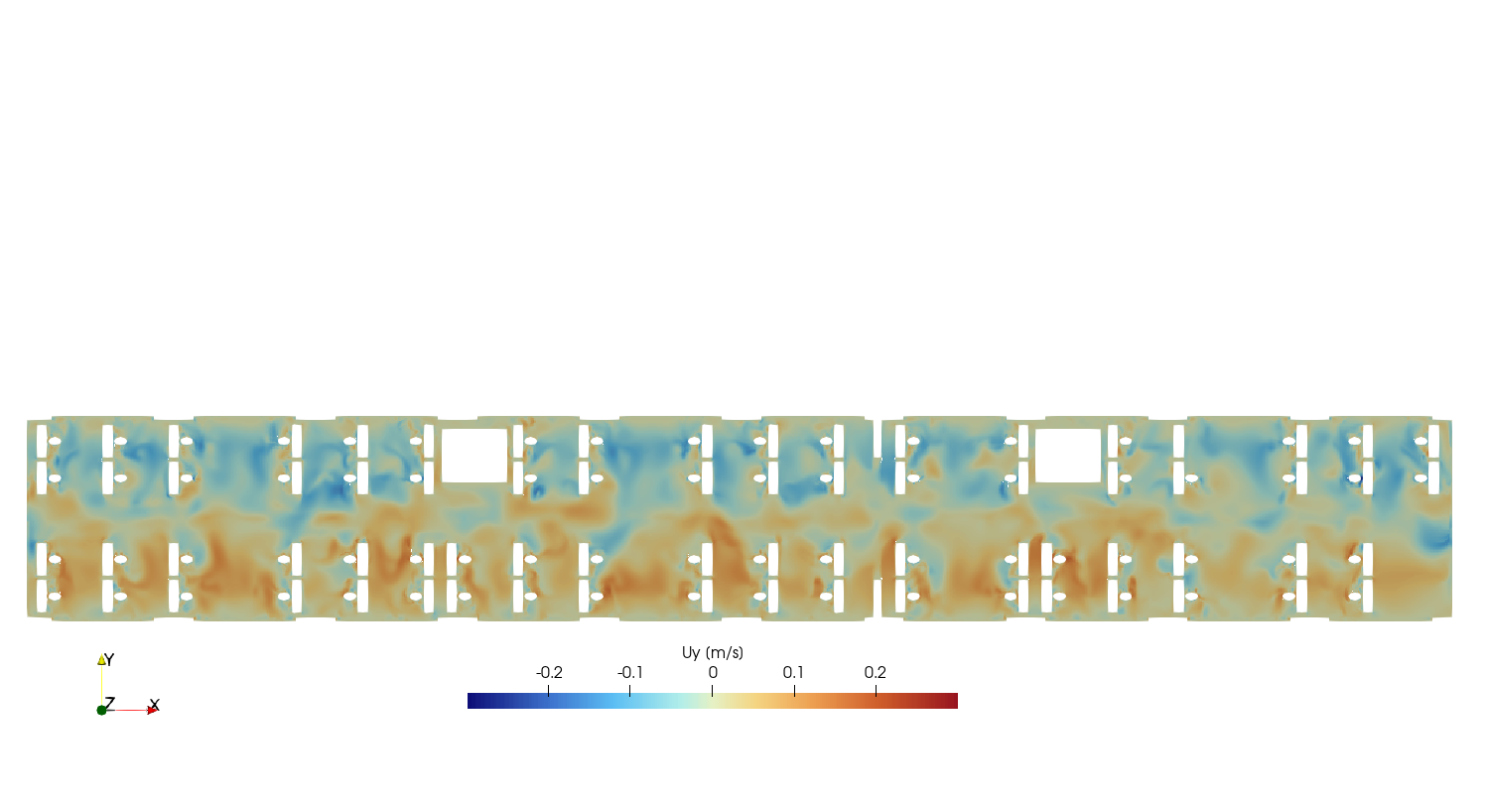}
    \caption{Split structure of $U_{y}$ in a slice at head-height through the complete compartment. Airflow is directed from the windows towards the aisle for all rows.}
    \label{fig:znormal_Uy}
\end{figure}

\begin{figure}
	\centering
	\includegraphics[width=\textwidth]{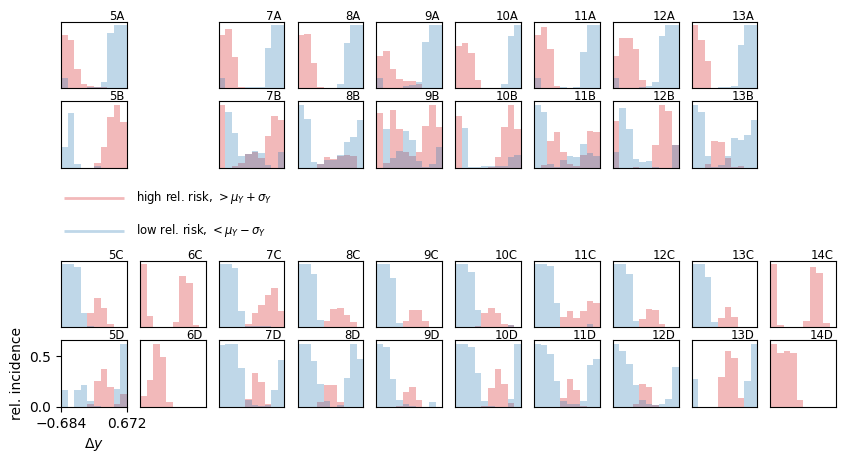}
	\caption{Input distributions of $\Delta y$ conditionally sampled conditioned on high and low resulting risks obtained from QMC. The sampling threshold is set at $\mu_{Y}\pm \sigma_{Y}$. Clear separation of the input distributions can be observed, consistent with a sensitive parameter of the risk model.}\label{fig:conditional_y_distribution}
\end{figure}

The high sensitivity of the average risk to offsets along the $x$ and $y$ axes is related to these circulations in between the rows of passengers.
Figure \ref{fig:znormal_Uy} shows the velocity component $U_{y}$ in a cross-section of the compartment at head level.
While the general flow structure of air moving from the windows toward the aisle, as shown in detail above for row 8, repeats for all other rows, the mean flow field exhibits many smaller structures, resulting in spatial variability in the magnitude of transport along $y$.
In the case of the dominating, directly adjacent source, $y$ offsets of the receiver, corresponding to shifting the breathing zone directly towards or away from the source, have a significant impact on the risk calculated for this critical case, and, consequently, on the overall average risk.
This effect is illustrated by figure \ref{fig:conditional_y_distribution}, showing the distributions of the QMC input values of $\Delta y$ conditionally sampled on the resulting risks according to a threshold of $\pm \sigma_{Y}$ around the seat-specific mean.
For seats where the lateral offset is the leading sensitive parameter, the distributions corresponding to high and low risks are clearly separated.
In these cases, pairs of neighboring seats, such as 5A and 5B, tend to show mirrored distributions since the lateral offset direction towards the neighbor alternates.

Similarly, shifts fore and aft along the $x$ axis serve to move the breathing zone out of the plane of the particle source, thereby also potentially avoiding most of the aerosol.
However, a competing effect occurs when the infectious passenger is seated directly in front or behind the subject, as $x$ offsets would then lead to an increased particle dose.

\section{Conclusions}
\label{sec:conclusion}
This investigation examines the impact of passenger head movement on the predicted average infection risk from airborne pathogens.
As a fully dynamic simulation of moving passengers is not feasible at reasonable computational cost, the movement was parametrized instead by offsets along three translational and two rotational axes.
These offsets were then modeled by sampling from normal distributions around an idealized central resting position.
The breathing zone positions and orientations were thus treated as input uncertainties affecting the probability distribution of the average infection risk for each seat.
The comparatively low computational cost of evaluating the breathing zones and the resulting infection risk enabled the efficient calculation of the output distributions via a pseudo-random quasi-Monte-Carlo (QMC) simulation.

The investigation showed that for the CFD-based forward infection risk prediction model, the average risk predicted for each passenger is highly sensitive to the exact location of their breathing zone.
Over the range of inputs, deviations by more than a factor of two from the mean were observed in the individual risks.
Including the variation in breathing zone position resulting from natural posture changes in this way adds another potentially important dimension to the model when informing design and policy decisions.
Comparing the QMC results with estimates obtained using local derivative-based methods showed that the latter are limited in their ability to estimate the statistical properties of the output distributions.
The dependence of the average risk on the breathing zone offsets is sufficiently nonlinear as to frustrate the inherently linear first-order second moment method, resulting in large errors in the predicted mean risks.
Using the second-order third moment method improves the prediction of the mean risk, reducing the error to below $10\,\%$.

When only considering the overall infection risk of all passengers collectively, the additional averaging allows for partial cancellation of over- and underestimated risk, thus smoothing the differences and reducing the relative error.
However, both methods fail to provide satisfactory estimates of the variance of the output distributions and are thus insufficient for more sophisticated risk assessment approaches that consider more than just mean risk.

To quantify the relative contributions of the separate inputs to the output variance, sensitivity measures were calculated in the form of the first-order and total Sobol' indices by fitting a polynomial chaos expansion (PCE) to the QMC data.
Analysis of these indices reveals that the predicted average infection risk is most sensitive to the translational offsets along the fore/aft ($x$) and the lateral ($y$) axes.
This sensitivity stems from the row-wise organization of quasi-two-dimensional flow structures in the form of counter-rotating circulations, which are responsible for transporting aerosols between adjacent seats.
Changes in the position of the breathing zone relative to the centroid of the exhaled particle cloud along the $x$ or $y$ axes therefore lead to significantly altered exposure levels and consequently infection risks.
Since the analysis of the relative contributions of each seat to the risk of each other seat generally shows dominance of nearby seats in general and of the immediate neighbor in particular, this effect translates to high sensitivity of the average risk as well.

As with the statistical moments, the feasibility of using derivative-based estimates for sensitivity analysis was explored by calculating the sigma-normalized derivatives as sensitivity indices.
Comparing these results to the total Sobol' index as a measure of total sensitivity to the input in question revealed limited agreement regarding the hierarchy of sensitivities.
Additionally, no significant advantage was found for the higher-order method.
In the case of infection risk prediction as presented here, estimating the output variance and consequently the decomposition of output variance into contributions of the different inputs, relies on knowledge of the dependencies across the entire range of the inputs.
Therefore, local approaches are inherently disadvantaged compared to global approaches, such as extracting Sobol' indices from PCE.
Complex three-dimensional flow fields and the resulting particle trajectories create irregular dependencies between the breathing zone position and orientation and the resulting average infection risk.
For this reason, global variance-based approaches are the preferred method for investigating the sensitivities in infection risk in passenger compartments.

The results presented here demonstrate the potential of incorporating behavior-based uncertainties into infection risk predictions involving airborne pathogens.
Exploiting the sequential pipeline of the forward risk prediction model allows results from the comparatively expensive flow and particle transport simulations to inform a large number of independent and inexpensive risk calculations.
This enables sampling-based approaches such as (quasi-) Monte-Carlo simulations.
Global, variance-based sensitivity analysis of the resulting risk distributions is a valuable screening tool for identifying possible technical and operational interventions to mitigate airborne transmission risk. 
Thus, the amount of actionable information that can be extracted from CFD-based particle trajectories via the infection risk model can be significantly increased at comparatively negligible additional computational cost.
Future work should aim at finding comparable methods to include spatial variations of the initial aerosol cloud, to provide the source-position complement to the receiver-focused results presented here.

\section*{Acknowledgments}
\begin{tabular}{ll}
	\includegraphics[width=0.22\textwidth]{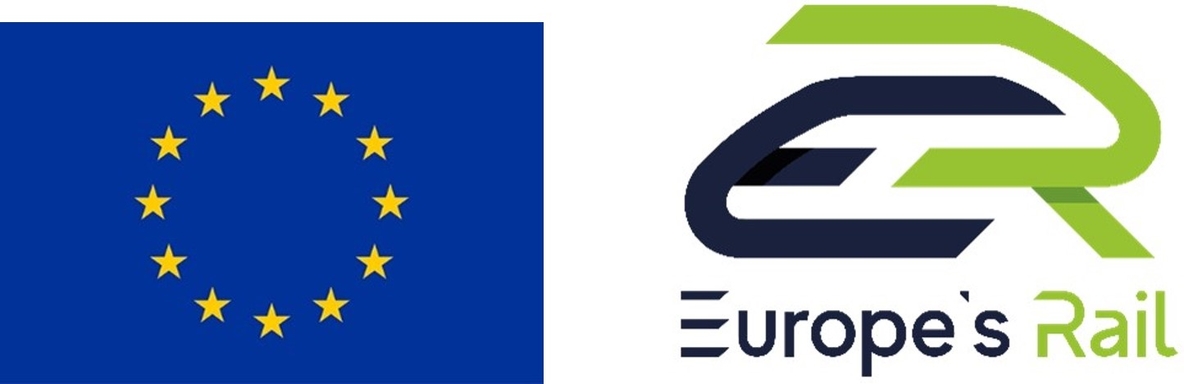} & \parbox{0.72\textwidth}{%
    Partially funded by the European Union. Views and opinions expressed are however those of the author(s) only and do not necessarily reflect those of the European Union or the Europe’s Rail Joint Undertaking. Neither the European Union nor the	granting authority can be held responsible for them. The project FP4-Rail4EARTH is supported by the Europe’s Rail Joint Undertaking and its members.%
    }\\
~\\
\end{tabular}
The authors gratefully acknowledge the scientific support and HPC resources provided by the German Aerospace Center (DLR). The HPC system CARO is partially funded by "Ministry of Science and Culture of Lower Saxony" and " Federal Ministry of Research, Technology and Space".

\clearpage

\bibliographystyle{unsrtnat}
\bibliography{bibliography}

\clearpage

\begin{appendices}
\section{Full train results}
\label{sec:app:full_results}
	\begin{figure}[h]
		\centering
		\includegraphics[width=0.8\textheight, angle=90, origin=c]{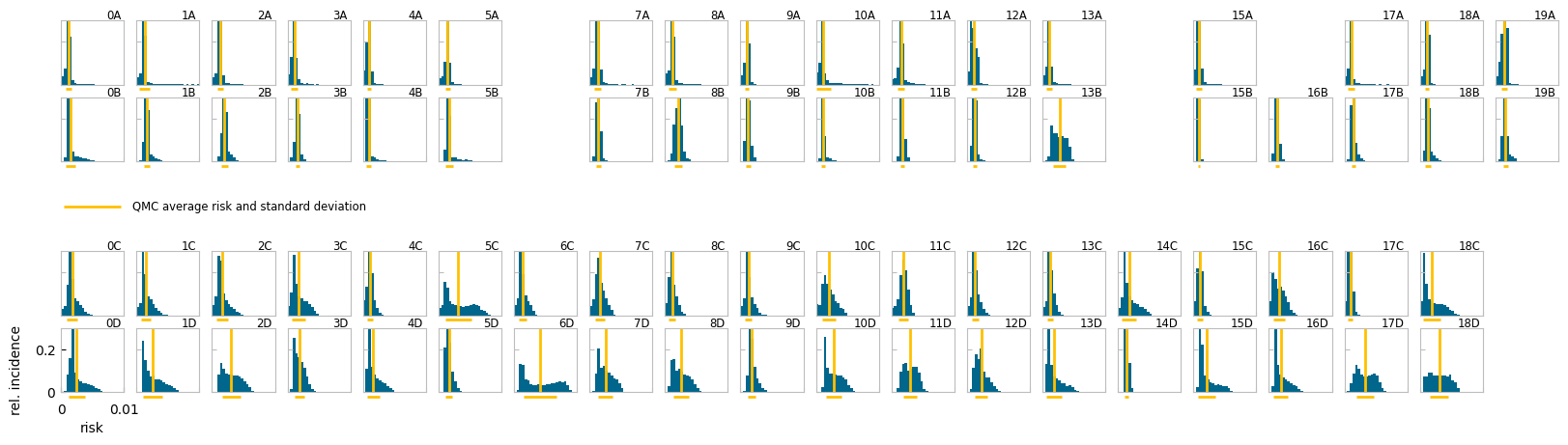}\hspace*{2cm}
		\includegraphics[width=0.8\textheight, angle=90, origin=c]{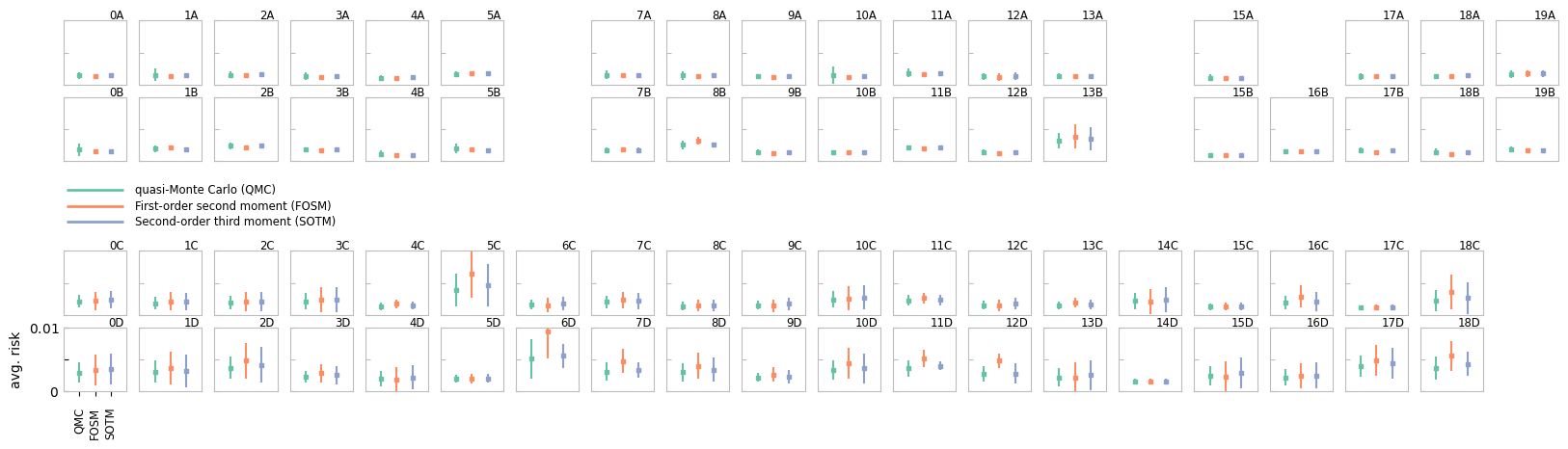}
		\caption{Left: Distributions of the mean infection risk for each passenger, averaged over all possible source positions, as produced by QMC simulations of $N=4096$ samples.\\
		Right: Estimates for the mean $\mu_{Y}$ and standard deviation $\sigma_{Y}$ of the resulting risk distribution as computed by QMC, FOSM, and SOTM.  }\label{fig:seat_risks_full}
	\end{figure}

	\begin{figure}
		\centering
		\includegraphics[width=0.8\textheight, angle=90, origin=c]{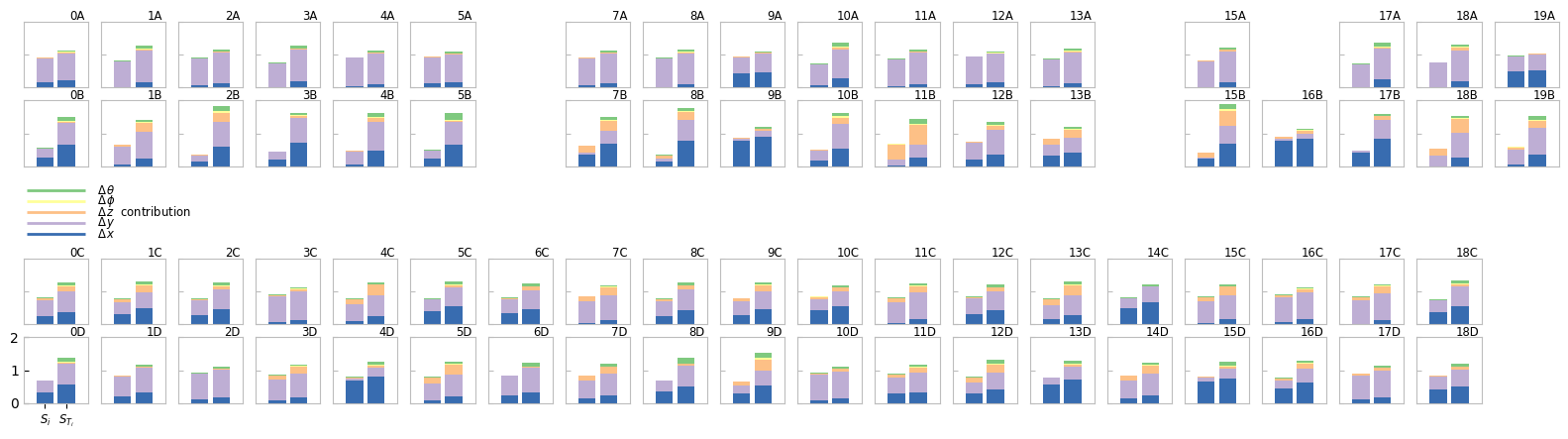}\hspace*{2cm}
		\includegraphics[width=0.8\textheight, angle=90, origin=c]{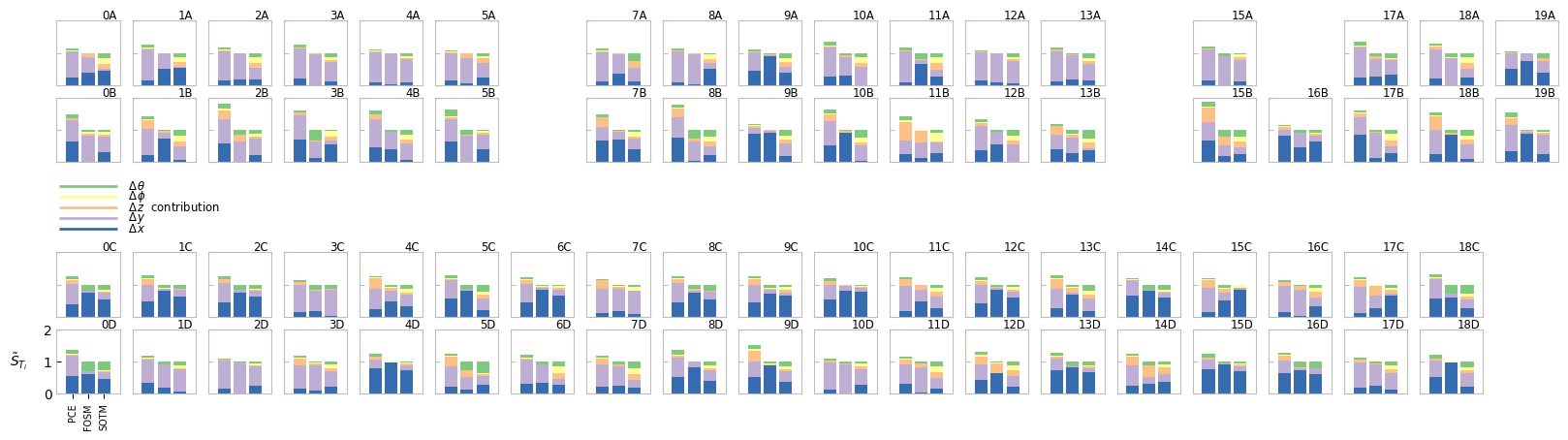}
		\caption{Left: First-order ($S_{i}$, left columns) and total indices ($S_{T_{i}}$, right columns) of the five model inputs for all passengers, computed from PCE of the QMC results. The respective height within the stacked bar corresponds to the value of the index in question.\\
		Right: Total sensitivity indices $S_{T_{i}}$ of the position parameters for the passengers seated in rows 5 through 14 with respect to the infection risk averaged over all possible source locations as derived from PCE compared to sigma-normalized derivatives $S_{\sigma_{i}}$ obtained from FOSM and SOTM.}\label{fig:sensitivities_full}
	\end{figure}

	\begin{figure}
		\centering
		\includegraphics{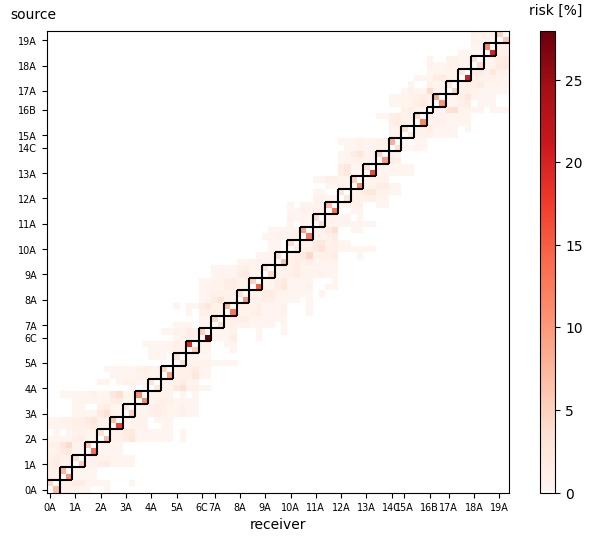}
		\caption{Average risk matrix obtained from QMC. Shown are the infection risks at a given receiver positions posed by a specific source positions for all possible combinations of seats, with the risk magnitude represented by the color of the corresponding matrix entry. The black outlines group directly neighboring seats together.}\label{fig:risk_matrix_full}
	\end{figure}

\end{appendices}

\end{document}